\documentclass[fleqn,usenatbib]{mnras}

\usepackage[T1]{fontenc}

\DeclareRobustCommand{\VAN}[3]{#2}
\let\VANthebibliography\thebibliography
\def\thebibliography{\DeclareRobustCommand{\VAN}[3]{##3}\VANthebibliography}

\usepackage{graphicx}	
\usepackage{amsmath}	
\usepackage{amssymb}	
\usepackage{newtxtext,newtxmath}

\newcommand{\kms}       {\mbox{${\rm km}\,{\rm s}^{-1}$}}

\newcommand{\kmskpc}    {\mbox{${\rm km}\,{\rm s}^{-1}\,{\rm kpc}^{-1}$}}

\newcommand{\B}[1]      {{\boldsymbol{#1}}}
\newcommand{\fxv}       {f({\bf x},\,{\bf v},\,t)}
\newcommand{\fxvp}       {f({\bf x}',\,{\bf v}',\,t)}

\title[GP Model for the Local Stellar Velocity Field]{Gaussian Process Model for the Local Stellar Velocity Field from {\it Gaia} Data Release 2}

\author[P. Nelson et al.]{
Patrick Nelson$^{1}$ \thanks{E-mail: patrick.nelson@queensu.ca} and
Lawrence M. Widrow$^{1}$\thanks{E-mail: widrow@queensu.ca}\\
$^{1}$Department of Physics, Engineering Physics, and Astronomy, Queen’s University,  Kingston, K7L 3N6, Canada\\}

\date{Accepted XXX. Received YYY; in original form ZZZ}

\pubyear{2022}

\begin{document}
\label{firstpage}
\pagerange{\pageref{firstpage}--\pageref{lastpage}}
\maketitle

\begin{abstract}
We model the local stellar velocity field using position and velocity measurements for 4M stars from the second data release of {\it Gaia}. We determine the components of the mean or bulk velocity in $\sim 27k$ spatially-defined bins. Our assumption is that these quantities constitute a Gaussian process where the correlation between the bulk velocity at different locations is described by a simple covariance function or kernel. We use a sparse Gaussian process algorithm based on inducing points to construct a non-parametric, smooth, and differentiable model for the underlying velocity field. We estimate the Oort constants $A$, $B$, $C$, and $K$ and find values in excellent agreement with previous results. Maps of the velocity field within $2\,{\rm kpc}$ of the Sun reveal complicated substructures, which provide clear evidence that the local disk is in a state of disequilibrium. We present the first 3D map of the divergence of the stellar velocity field and identify regions of the disk that may be undergoing compression and rarefaction.
\end{abstract}

\begin{keywords}
Galaxy: kinematics and dynamics -- Galaxy: structure 
\end{keywords}



\section{Introduction} \label{sec:Introduction}
A common strategy for understanding the dynamics of the Milky way is to assume that phenomena such as the bar, spiral arms, and warp can be understood as perturbative departures from an equilibrium state (See, for example, \citet{binney2013,sellwood2013}). A further assumption is that stars orbit in the mean gravitational field of gas, dark matter, and the other stars. The stellar components of the Galaxy are then described by a phase space distribution function (DF), $f({\bf x},\,{\bf v},t)$, which obeys the collisionless Boltzmann equation (CBE) coupled to gravity via Poisson's equation. The equilibrium/perturbation split then carries over to the DF and gravitational potential $\Psi$. For a disk galaxy such as the Milky Way, the equilibrium state exhibits symmetry about both the its midplane and rotation axis. Any departure from these symmetries therefore signals a departure from equilibrium.

The second {\it Gaia} data release ({\it Gaia} DR2, \citet{Gaia2018a}) includes measurements of positions and velocities for over seven million stars thereby vastly increasing the number of stars for which all phase space coordinates are known. This data provides an estimate of the present-day stellar DF near the Sun in the sense that each star can be represented as a phase space probability distribution function $P_i$ that reflects observational uncertainties in its position and velocity:
\begin{equation}
    \fxv = \sum_i P_i\left ({\bf x}-{\bf x}_i,{\bf v}-{\bf v}_i\right )~.\end{equation}\label{eq:DF}
In the limit of perfect observations, $P_i$ reduces to a six-dimensional delta function. In general, interpretation of the DF involves the construction of a set of observables 

\begin{equation}
    O = \int d^3x' d^3 v' \fxvp {\cal F}({\bf x'},\,{\bf v'})
\end{equation}

\noindent where ${\cal F}$ is designed to pick out particular features of the DF and also account for the selection function of the survey. For example, an estimate of the number density $n({\bf x})$ is obtained by setting ${\cal F}={\cal W}({\bf x}-{\bf x'})$ where ${\cal W}$ is a localized window function:
\begin{equation}
    n({\bf x}) = \int d^3x' d^3v' \fxvp {\cal W}({\bf x}-{\bf x'})~.
    \end{equation}
In this paper, we focus on the first velocity moment of the DF:
\begin{equation}
    {\bf V}({\bf x}) = n^{-1}\int d^3x' d^3v' \fxvp {\cal W}({\bf x}-{\bf x'})\,{\bf v}~.
    \end{equation}

\noindent For a system in equilibrium, ${\bf V}$ is a function of Galactocentric radius $R$ and distance from the midplane $|z|$.

Studies of the bulk velocity in pre-{\it Gaia} surveys such as the Sloan Extension for Galactic Understanding and Exploration (SEGUE) \citep{yanny2009}, the Radial Velocity Experiment (RAVE) \citep{steinmetz2006}, and  Large Sky Area Multi-Object Fiber Spectroscopic Telescope (LAMOST) \citep{cui2012} revealed bulk motions than signaled a Galaxy in disequilibrium. For example, \citet{widrow2012} determined the bulk vertical velocity $V_Z$ and its dispersion $\sigma_z$ as functions of $z$ for SEGUE stars and found that they showed features asymmetric about $z=0$. \citet{Williams_2013} determined the full 3D velocity field from some 70k red clump stars in the RAVE survey and found vertical motions that varied with Galactocentric radius $R$ as well as evidence for radial flows. Similarly, \citet{carlin2013} and \citet{pearl2017} used LAMOST spectroscopic velocities with proper motions from the PPMXL catalog \citep{roeser2010} for 400k stars to map out both vertical and radial bulk motions. Finally, \citet{Schoenrich2019} and \citet{friske2019} found a wavelike pattern in the mean vertical velocity as a function of angular momentum about the spin axis of the Galaxy, which served as a proxy for Galactocentric radius.

Unfortunately, a clear picture of bulk motions near the Sun from these and other studies never emerged. The surveys each had their own complicated selection functions, as illustrated in Figure 1 of \citet{carlin2013}. Furthermore, systematic distance errors could masquerade as velocity-space substructure \citep{Schoenrich2012, carrillo2018}. These problems were alleviated by \citet{Katz2019} who presented a series of velocity field maps using data from {\it Gaia} DR2 and straight-foward binning procedures. For example, they constructed face-on maps by first dividing the sample into vertical slabs of width $\Delta z = 400\,{\rm pc}$ and then computing the mean velocity in $xy$-cells of $200\,{\rm pc}$ by $200\,{\rm pc}$. Similar binning procedures were used to construct maps of the velocity field in the $R-z$ and $R-\phi$ planes.

The {\it Gaia} kinematic maps are simple and intuitive though the connection with theory, say through the CBE, requires further analysis. In particular, the mean velocities in the {\it Gaia} maps differ from the moments that appear in the continuity and Jeans equations since the former are averaged over cells of a finite size that must strike a balance between improved spatial resolution and reduced statistical fluctuations. The situation is only exacerbated when one goes to compute gradients of the velocity field.

In this paper, we take a different approach based on Gaussian process (GP) regression, a Bayesian, (almost) nonparametric approach to fitting data from the field of machine learning (See, for example, \citet{Rasmussen_book_2006}). GP regression has been used extensively in data science and engineering though only recently has it been applied to problems in astrophysics. An example that has some similarities with the problem at hand is the construction of dust and extinction maps in the Galaxy (see, for example, \citet{sale2018}). (For a different approach for constructing smooth maps of streaming motions, see \citet{khanna2022}.)

Our provisional starting point is the assumption that the velocities of stars in the Galaxy constitute, to a good approximation, a Gaussian process. Formally, a Gaussian process is a collection of random variables such that any subset of these variables is Gaussian. Stellar velocities would seem to be a good candidate for a GP since the velocity distributions for the differentstellar components (thin and thick disks, stellar halo) are well-described by anisotropic Maxwellians \citep{binney2008}. 

The fundamental object in GP regression is the covariance matrix, which describes correlations in the output variables for different values of the inputs. In our case, the outputs are the components of the velocity field and the inputs are different positions within the Galaxy. The covariance matrix is constructed from a kernel function, which depends of a set of hyperparameters that control the strength and scale length of correlations in the velocity field. Thus, although the model for the velocity field itself is nonparametric, the covariance function is parametric. The driver in GP regression is optimization of the likelihood function, defined below, over the space of hyperparameters. Note that GP regression is different from kernel smoothing or kernel density estimation, where one produces a smooth model by convolving the data with a window function. In GP regression, we use the data to select a model from the space of nonparametric functions in the GP prior.

To evaluate the GP likelihood function, one must invert an $N$ by $N$ covariance matrix where $N$ is the number of observations. This calculation is an $O(N^3)$ operation that requires $O(N^2)$ 
rapid-access memory (RAM) capabilities. Clearly, it is unfeasible to apply GP regression directly to the {\it Gaia} data, where $N$ in DR2 already exceeds $10^6$. In this work, we apply a two-fold strategy for handling the large-$N$ bottleneck. First, we bin stars into cells of size $(\Delta x,\,\Delta y,\,\Delta z) = (125,\,125,\,\,50)\, {\rm pc}$. The factor of $2.5$ difference in $\Delta z$ is meant to account for the difference between the radial scale length and the vertical scale height of the disk. Note that the width of the bins in $z$ are a factor of $8$ smaller than those used in constructing the \citet{Katz2019} face-on maps. Our binning procedure leads to roughly $27k$ cells above our threshold of $20$ stars. We then use a sparse GP algorithm that deploys $M$ inducing points to approximate the likelihood function. This algorithm reduces the computational complexity to $O(NM^2)$ and the RAM requirements to $O(NM)$.

Our GP analysis leads to a single model from which other properties of the velocity field can be derived. By contrast \citet{Katz2019} use different binning strategies to explore different facets of the velocity field. Moreover, the {\it Gaia} maps are somewhere between a projection, where one integrates out one spatial dimension, and a two-dimensional slice, where one takes a narrow range in one of the dimensions to get the velocity field on a two-dimensional surface. By contrast, our GP model can be queried to give the inferred velocity field with uncertainties at any point in the sample region. Furthermore, since the model is differentiable, We can use it to infer quantities such as the Oort constants at the position of the Sun. In addition, derivatives of ${\bf V}$ can be used to make statements about the dynamical state of the Galaxy. This feature allows us to map out variations in Oort functions in the vicinity of the Sun. It also allows us to construct a 3D map of the divergence, which is proportional to the total time derivative of the stellar number density. In so doing, we are able to identify regions near the Sun where the number density of stars may be undergoing compression and rarefaction.

An outline of our paper is as follows. In Section \ref{sec:Preliminaries} we describe the sample used in our analysis as well as our binning strategy. In Section \ref{sec:GPR} we provide a brief introduction to Gaussian processes and outline our sparse GP algorithm. We also summarize results of tests performed with mock data. The results of our analysis for {\it Gaia} DR2 are discussed in Sections \ref{sec:Results} and \ref{sec:Velocity-Gradient}. We present estimates for the generalized Oort constants and maps of the velocity field as well as a three-dimensional map of the divergence of the velocity field. We discuss possible extensions of this work in Section \ref{sec:Discussion} and provide a brief summary of our results and some concluding remarks in Section \ref{sec:Conclusion}.

\section{Preliminaries} \label{sec:Preliminaries}
\subsection{6D phase space catalog} \label{sec:6D-catalog}
The {\it Gaia} DR2 Radial Velocity survey contains 6D phase-space measurements for over 7 million stars \citep{Gaia2018a,Katz2019}. In this work, we use the {\it gaiaRVdelpepsdelsp43} catalog, which was constructed to correct for systematic errors in the {\it Gaia} parallaxes and uncertainties \citep{Schoenrich2019}\footnote{See https://doi.org/10.5281/zenodo.2557803}.

Following \citet{Katz2019} we use both Cartesian coordinates $(X,Y,Z,U,V,W)$ and Galactic cylindrical coordinates $(R,\phi,Z,V_R,V_\phi,V_Z)$.
The Cartesian coordinate system has the Galactic centre at the origin and the Sun on the $-X$-axis; $U,\,V$ and $W$ are the usual velocity components where positive values at the position of the Sun indicate motion toward the Galactic centre, the direction of Galactic rotation, and the North Galactic Pole, respectively.
We use $8.27\,{\rm kpc}$ for the Sun's distance to the Galactic centre and $20\,{\rm pc}$ for its distance to the Galactic midplane (see \citet{Schoenrich2019} and references therein) so that $(X_\odot,\,Y_\odot,\,Z_\odot) = (-8.27,\,0,\,0.02)\,{\rm kpc}$. We further set $\left (U_\odot,\,V_\odot,\,W_\odot\right ) = \left (11.1,\,250,\,7.24\right )\,{\rm km}\,{\rm s}^{-1}$ \citep{Schoenrich2012}. Our cylindrical coordinate system also has the Galactic centre at the origin, the Sun at $\phi=0^\circ$, and $\phi$ increasing in the direction of Galactic rotation. This system is left-handed in the sense that increasing $\phi$ corresponds to the direction toward the South rather than North Galactic pole. We note that at the position of the Sun $V_R = -U$ and $V_\phi=V$; $V_Z = W$.

We calculate the $U,\,V, W$ velocity components and their errors from astrometric observations using the method described in \citet{johnson1987}. We then convert to $V_R$, $V_\phi$ and $V_Z$.
The velocity components, but not their errors, were provided with the {\it gaiaRVdelpepsdelsp43} catalog. Note that in that catalog, the Galactocentric cylindrical velocity components were given as $(U_g,\, V_g,\, W_g) = (-V_R,\,V_\phi,\,V_z)$ and the Cartesian coordinates were given as $(x,\,y,\,z) = (-X,\,Y,\,Z)$. 

We apply the following restrictions to the sample:

\begin{itemize}
    \item color: $G_{bp} - G_{rp} < 1.5$
    \item magnitude: $G<14.5$ and $G_{bp},\,G,\,G_{rp} > 0$
    \item parallax signal to noise: $p/\sigma_p > 4$
    \item parallax uncertainty: $\sigma_p < 0.1\,{\rm mas}$ with $\sigma_p$ given by the {\it Gaia} pipeline
    \item excess $B-R$ flux: $1.172 < E_{bprp} < 1.3$
\end{itemize}

\noindent These quality cuts are recommended in \citet{Schoenrich2019} to ensure minimal systematic biases in derived kinematic quantities. They leave 4584106 stars from the original catalog of 6606247. Finally, we apply the following kinematic cuts to remove high-velocity outliers:

\begin{itemize}
    \item proper motion: $\mu_\alpha,\,\mu_\delta < 400\,{\rm mas\,yr^{-1}}$
    \item proper motion error: $\epsilon_{\mu_\alpha},\,\epsilon_{\mu_\delta} < 20\,{\rm mas}\,yr^{-1}$
    \item Galactocentric speed: $|{\bf V}| < 600\,{\rm km\,s}^{-1}$
\end{itemize}

\noindent These cuts are similar to those implemented by \citet{Williams_2013} with the modification that our velocity cut is in terms of the speed in the local frame of rest whereas they apply a cut on the radial velocity. These cuts eliminate only about 400 stars. 

\subsection{Binning} \label{sec:Binning}
As discussed above, computation time and RAM requirements make it unfeasible to apply GP regression directly to data sets with much more than $10^4$ entries. For this reason, we bin stars so that input data for our GP analysis are the mean velocity components in cells. In a sense, velocity components averaged over stars in a cell are closer to  Gaussian process than the stellar velocities themselves. Stars in the region near the Sun can come from the thin disk, the thick disk, or the stellar halo. The velocity distributions for these components are roughly Maxwellian. Thus, the stellar velocities are drawn from what might be better described as a mixture of Gaussians. On the other hand, the average velocity of some large number of stars in a cell will be approximately Gaussian due to the central limit theorem. 

We set up a Cartesian grid of cells with size $(\Delta x,\,\Delta y,\,\Delta z)=(125,\,125,\,\,50)\,{\rm pc}$ for the region $4 < X ({\rm kpc})  < 12$, $-4 < Y ({\rm kpc}) < 4$, and $-2 < Z ({\rm kpc})< 2$ and keep only those cells with more than 20 stars. Mean values and uncertainties for the $V_R, \,V_\phi$, and $V_Z$ components are then calculated by a standard least squares algorithm. In the end, we are left with mean velocities for 27305 cells representing the observations of 3972825 stars. 

\section{Gaussian Process Regression} \label{sec:GPR}

\subsection{Overview of Gaussian processes}

We begin with a brief review of GP regression. More thorough discussions can be found in numerous resources such as the excellent book by \citep{Rasmussen_book_2006}. Suppose we have observations of a real scalar process $f$ such that

\begin{equation}
    y_i = f({\bf x}_i) + \epsilon_i~~~~~i=\{1,\,2,\,\dots N\}
\end{equation}

\noindent where ${\bf x}_i$ is the input vector for the $i$'th observation, $y_i$ is the scalar output and $\epsilon_i$ is additive noise for that observation. For the case at hand, ${\bf x}_i = (X_i,\,Y_i,\,Z_i)$ is the position vector of the $i$'th cell and $f$ is $V_R$, $V_\phi$, or $V_Z$. If $f$ is a Gaussian Process then it is completely specified by a mean function $m({\bf x})$ and covariance function $k({\bf x},{\bf x}')$ such that

\begin{align}
    m({\bf x}) & = \mathbb{E}\left [f({\bf x})\right ]
\\
    k({\bf x},\,{\bf x}') & = \mathbb{E}\left [
    \left (f({\bf x})-m({\bf x})\right )
        \left (f({\bf x}')-m({\bf x}')\right )
    \right ]
    \end{align}

\noindent where $\mathbb{E}$ denotes expectation value. Furthermore, the probability distribution function of ${\mathbf f} \equiv \left [f_1,\,\dots,f({\bf x}_n\right ]$ is given by
\begin{equation}\label{eq:prior}
       P(\mathbf{f}\,\lvert\,\mathbf{X}) = \mathcal{N}(\mathbf{f}\,\lvert\, \boldsymbol\mu, \mathbf{K})\,
\end{equation}
where $\mathbf{X}$, $\boldsymbol\mu$, and $\mathbf{K}$ are aggregate vectors of the input vectors ${\bf x}_i$, the mean functions $m_i\equiv m({\bf x}_i)$, and the kernel functions
$K_{ij} \equiv k(\mathbf{x}_i,\mathbf{x}_j)$, respectively.
As usual, ${\cal N}$ denotes a multivariate Gaussian. Equation\,\ref{eq:prior} constitutes a Gaussian process prior on $f$. For simplicity, we assume $m=0$. 

The goal of GP regression is to infer $\mathbf{f}_*$ at new inputs ${\bf X}_*$ given the data $\left \{\mathbf{X},\,\mathbf{y}\right \}$.
If the noise $\epsilon$ is identical, independent, and Gaussian with dispersion $\sigma_n$, then the joint probability for $\mathbf{y}$ and $\mathbf{f}_*$ is itself Gaussian
\begin{equation}
P(\mathbf{y},\,\mathbf{f}_*|\mathbf{X},\,\mathbf{X}_*) =  \mathcal{N}\left(0,\, \begin{bmatrix}\mathbf{K} + \sigma_n^2 {I} & \mathbf{K}_* \\ \mathbf{K}_*^\mathsf{T} & \mathbf{K}_{**}\end{bmatrix}\right) \ ,
\end{equation}
where $\mathbf{K}\equiv K(\mathbf{X}, \mathbf{X})$, $\mathbf{K}_* \equiv K(\mathbf{X}, \mathbf{X}_*)$ and $\mathbf{K}_{**}\equiv K(\mathbf{X}_*, \mathbf{X}_*)$. The quantity of interest is the conditional probability
\begin{align}
    P(\mathbf{f}_*|\mathbf{y},\,\mathbf{X},\,\mathbf{X}_*)
    & = \frac{P(\mathbf{y},\,\mathbf{f}_*|\mathbf{X},\,\mathbf{X}_*)}{P(\mathbf{y})}\\
& = {\cal N}\left (\bar{\mathbf{f}}_*,\,C\right )
\end{align}
where 
\begin{equation}\label{eq:posterior}
    \bar{\mathbf{f}}_* = \mathbf{K}_*^T \left (\mathbf{K} + \sigma^2\mathbf{I}\right )^{-1} \mathbf{y}
\end{equation}
\noindent and 
\begin{equation}\label{eq:posteriorcov}
    \mathbf{C} = \mathbf{K}_{**} - \mathbf{K}_*^T
    \left (\mathbf{K} + \sigma^2\mathbf{I}\right )^{-1}\mathbf{K}_*
\end{equation}
\noindent (see \citet{Rasmussen_book_2006}, section 2.2 and appendix A).

\subsection{Kernel function}

A key ingredient of GP regression is the kernel or covariance function. Though there is considerable flexibility in choosing a kernel, for most problems it is constructed by taking sums and/or products of standard kernels, which themselves are constructed from elementary functions of the input variables.

In the present analysis, we allow for different kernels for $V_R$, $V_\phi$, and $V_Z$ since they are modeled as independent scalar functions. Each of the kernels uses a three-dimensional radial basis function (RBF) plus a term proportional to the identity matrix that accounts for unknown noise
\begin{equation}\label{eq:rbf}
    k({\bf x}_i,\,{\bf x}_j) = 
    \sigma_f^2 e^{-\tilde{r}_{ij}^2/2} + \sigma_n^2\delta_{ij}
    \end{equation}
\noindent where $\sigma_f^2$ is the signal variance and
\begin{equation}   
\tilde{r}_{ij} \equiv
    \frac{(X_i-X_j)^2}{l_x^2}
    +\frac{(Y_i-Y_j)^2}{l_y^2}
    + \frac{(Z_i-Z_j)^2}{l_z^2}
\end{equation}
is a dimensionless pseudo-distance between ${\bf x}_i$ and ${\bf x}_j$. 
The RBF kernel is positive definite, differentiable, and maximumal for ${\bf x}_i = {\bf x}_j$. These are all features that lead to realistic models for the bulk velocity field of the Galactic disk. Through trial and error, we find that the fit for $V_\phi$ is significantly improved by including in $k$ the product of two linear kernels. The linear kernel is given by
\begin{equation}
    k_{\rm lin}({\bf x},\,{\bf x}') = \sum_{i=x,y,z}\mu_i x_ix_i'~.
\end{equation}
\noindent It expands the space of priors on $f$ to include linear functions of the inputs ${\bf x}$. Unlike the RBF kernel, it is non-stationary in the sense that it depends on the absolute positions of the data points rather than the distance between pairs of data points. For $V_\phi$, we add
\begin{equation}
    k_{\rm lin2} = \left (\sum_{i=x,y,z}\mu_i x_ix_i'\right )\left (
    \sum_{j=x,y,z}\nu_j x_jx_i'\right )
\end{equation}
to the RBF and noise kernels in Equation \ref{eq:rbf}.

The kernels defined above depend on a number of free parameters commonly referred to as hyperparameters, since they parameterize the covariance function rather $f$ itself. The hyperparameters determine characteristics of functions in the prior of $f$. For example, $l_x$, $l_y$, and $l_z$ control the correlation lengths along the three coordinate axes. Proper choice of hyperparameters is crucial for obtaining a good model of the data. If the length scales are too small, then the model will tend to overfit the data, thereby attributing small-scale bumps and wiggles to $f$ rather than noise. Conversely, if the length scales are too large, then the model will tend to miss important features represented in the data.

In very simple problems, one can find suitable hyperparameters by trial and error as illustrated in Chapter 2 of \citet{Rasmussen_book_2006}. A principled approach, and one that is suitable for more complex problems, is to maximize the marginal likelihood function $p(\mathbf{y}|{\mathbf X},\theta)$ where
\begin{align}\label{eq:GPLike}
\log(p(\mathbf{y}|{\mathbf X},\theta) & = -\frac{n}{2}\log(2\pi)
 -\frac{1}{2} \log|K + \sigma_n^2I|\nonumber\\
 & - \frac{1}{2}\mathbf{y}^T(K+\sigma_n^2)^{-1}\mathbf{y}~.
\end{align}
Here, $\theta$ represents the hyperparameters.

\subsection{Sparse GP regression via inducing points}

In any optimization scheme, one must evaluate the marginal likelihood function in Equation\,\ref{eq:GPLike} a large number of times for different choices of the hyperparameters $\theta$. Each likelihood call involves  inversion of the $N\times N$ matrix $K$, which is an $O(N^3)$ operation that requires $O(N^2)$ RAM. This process becomes unfeasible for $N$ much greater than $10^4$. Fortunately, there are a number of algorithms that allow one to estimate the marginal likelihood using $M<N$ inputs. These algorithms, generally referred to as sparse GP regression, reduce the computational complexity to $O(NM^2)$ and the RAM requirement to $O(NM)$. In this work, we follow the method described in
 \citep{Bauer2017}. We denote the full covariance matrix ($\mathbf{K}$ in Equation\,\ref{eq:prior}) as $\mathbf{K}_{ff}$, the covariance matrix for the inducing points as $\mathbf{K}_{uu}$ and the cross covariance matrix as $\mathbf{K}_{fu}$ In sparse GP regression, one replaces $\mathbf{K}_{ff}$ in Equation\,\ref{eq:GPLike} with $\mathbf{Q}_{\rm ff}\equiv \mathbf{K}_{\rm fu}\mathbf{K}_{\rm uu}^{-1}\mathbf{K}_{\rm uf}$ and includes an additional term given by ${\rm Tr}(\mathbf{K}_{\rm ff}-\mathbf{Q}_{\rm ff})/2\sigma_n^2$
\citep{Titsias2009,Bauer2017}. Note that optimization requires that we compute the gradient of the marginal likelihood with respect to the hyperparameters.

We implement our GP algorithm using \textsc{GPy}, a Gaussian process package written in \textsc{Python} by the University of Sheffield machine learning group \citep{gpy2014}. We apply the sparse GP module with a heteroscedastic Gaussian likelihood, which allows us to incorporate uncertainties in the mean velocity components for individual cells. Optimization is carried out by applying the \textsc{VarDTC} inference method \citep{Titsias2009} using  the \textsc{GPy} optimization module and the \textsc{L-BFGS-B} algorithm from the software package \textsc{SciPy} \citep{Numerical_Optimization,2020SciPy}.

\subsection{Mock data tests} \label{sec:Mock-Data}

\begin{figure}
	\includegraphics[width=\columnwidth]{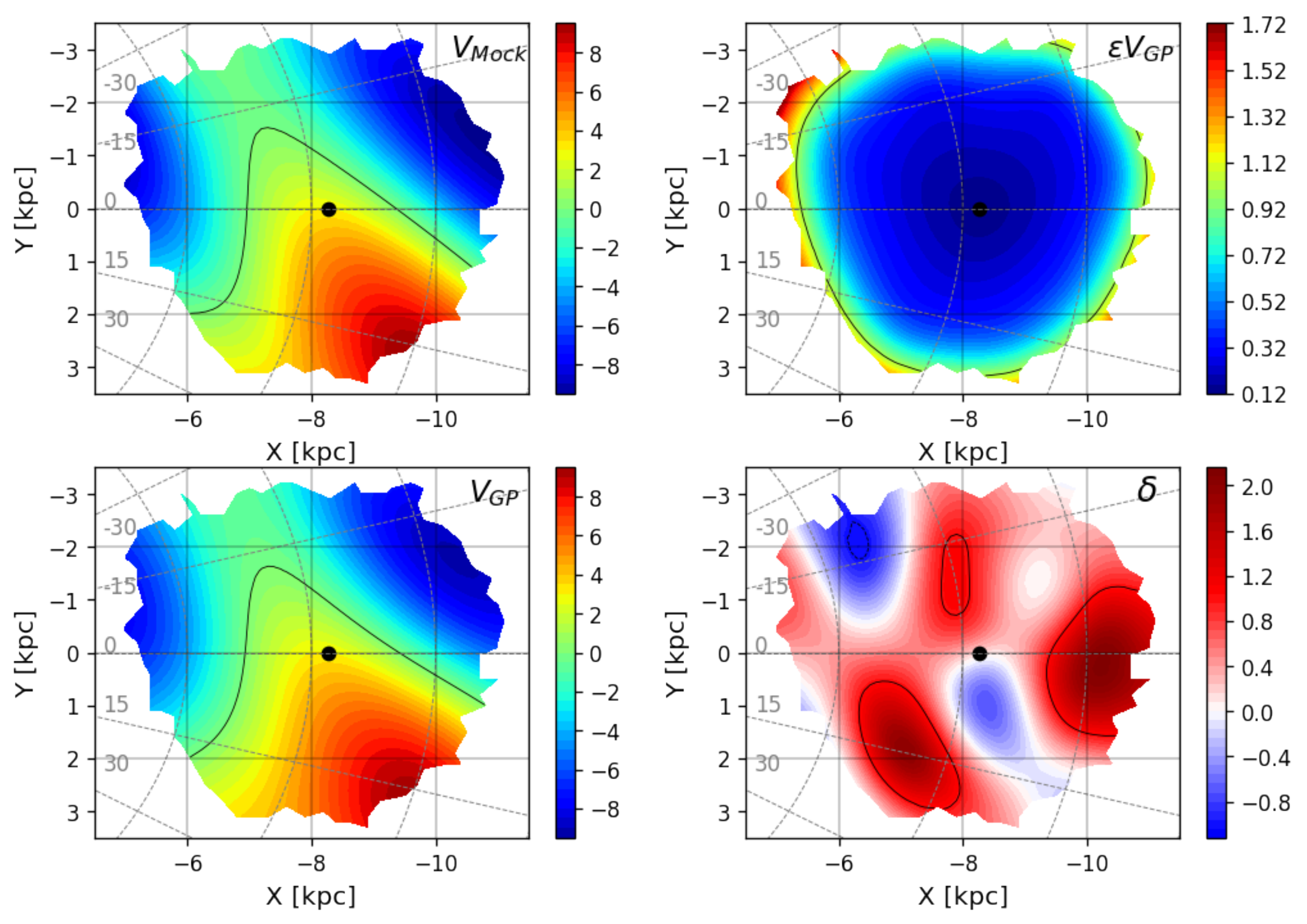}
    \caption{Results of a mock data test. Top left: the GP fit of the mock data. Top right: the confidence region of the model. Bottom left: the generating field for the mock data. Bottom right: the difference between the mock field and the model, scaled by the dispersion as calculated in the GP regression. Here and throughout, the black dot indicates the position of the Sun and velocities are given in $\rm km\,s^{-1}$.}
    \label{fig:Mock-Data}
\end{figure}

\begin{figure}
	\includegraphics[width=\columnwidth]{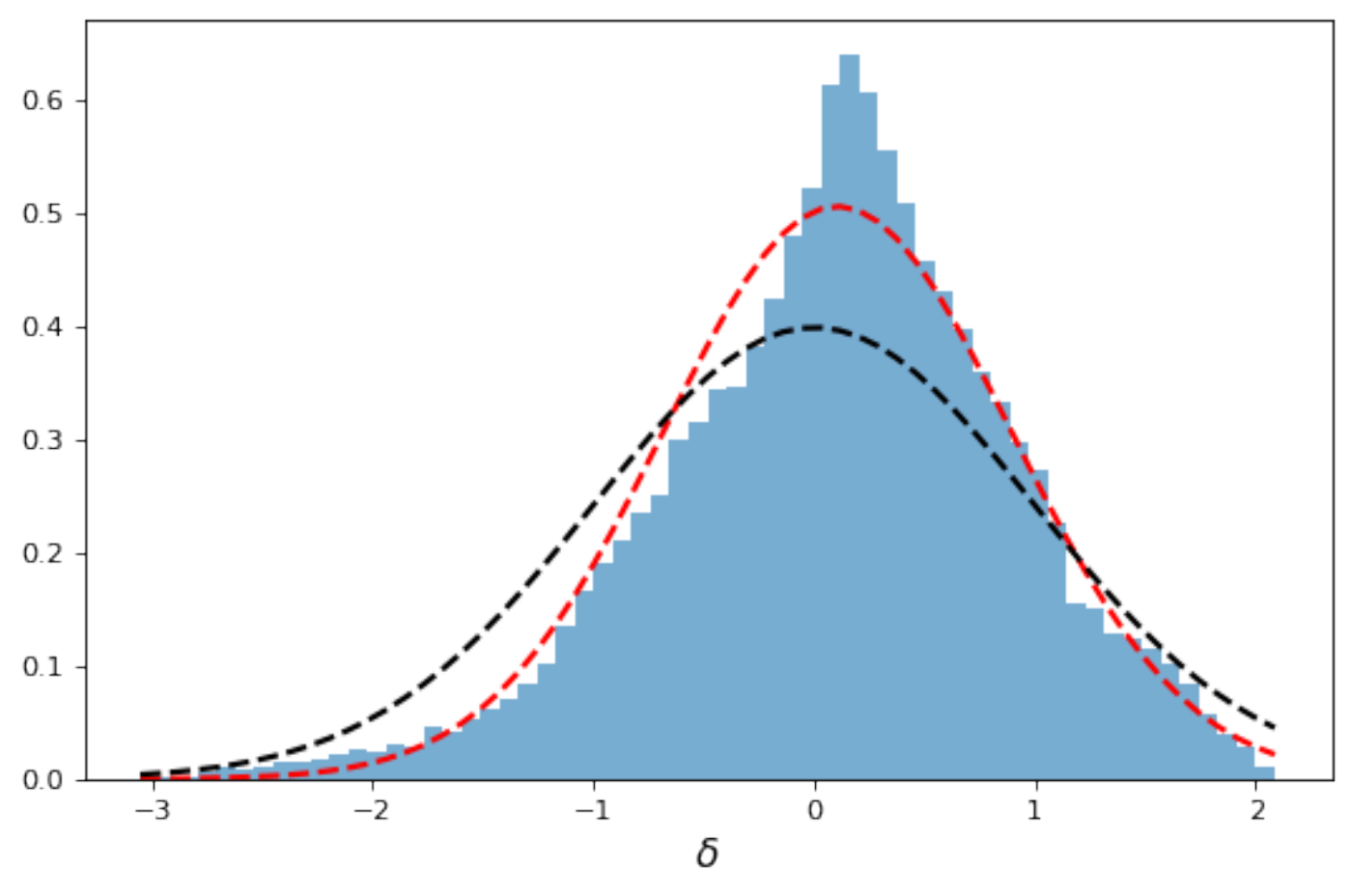}
    \caption{Histogram of the normalized residual  $\delta$ as defined in the text. The histogram includes residuals from all cells used in fitting the mock data and is itself normalized so that the area under the curve is unity. The black dashed line is the normal distribution ${\cal{N}}(0,1)$. The red dashed line is the normal distribution fit ${\cal{N}}(0.11,0.79)$.}
    \label{fig:Mock-Data-Hist}
\end{figure}

Before turning to {\it Gaia} DR2, we test our algorithm on a mock data sample. This sample is constructed by replacing measured velocities in the {\it gaiaRVdelpepsdelsp43} catalog with velocities drawn from an analytic function that is chosen so that the resulting maps are qualitatively similar to ones found with the real data. For brevity, we present results for a single generic component with velocity field. The analytic function is given by 
\begin{equation}
    V(X,Y,Z) = 2 \left (11\cos{g_1} - 3\cos{g_2} - 2\cos{g_3} - Z^2 -10 \right)
\end{equation}
\noindent where $g_1 =Z+(X-8)/6$, $g_2 =    0.52(Y-2)$, and $g_3 = X - 1.3Y - 11.3$.
The velocity field for $Z=0$ is shown in the upper left panel of Figure \ref{fig:Mock-Data}.

The mock data is analysed using the same procedure that will be applied to the real data. We first determine mean values for $V$ and errors about the mean for the $27k$ cells. We then optimize the likelihood function over the hyperparameters using the inducing point algorithm described above with $M=3000$ and the RBF plus noise kernel function. Armed with optimal hyperparameters, we determine the posterior of the model velocity field $\bar{V}$ using Equation\,\ref{eq:posterior}. The result for the $XY$-plane is shown in the lower left panel of Figure \ref{fig:Mock-Data}. We also determine the covariance matrix of the posterior using Equation \ref{eq:posteriorcov}. The square root of its diagonal elements provides an estimate of the dispersion, $\epsilon_V$, which is shown in the upper right panel of Figure \ref{fig:Mock-Data}. As expected, the dispersion rises sharply toward the edges of the region where we have data. Finally, we show the normalized residual $\delta\equiv (V-\bar{V})/\epsilon_V$ in the $XY$ plane. In Figure \ref{fig:Mock-Data-Hist}, we plot a histogram of $\delta$ over the 
sample volume. If the agreement between model and mock data were perfect, we would expect the histogram to be well-approximated by the normal distribution, ${\cal N}(0,1)$. Instead, we find that the histogram is fit by the Gaussian ${\cal N}(0.11,0.79)$. This indicates  that systematic errors are roughly a factor of 10 smaller than statistical ones and that uncertainties are over-estimated by $\sim 20\%$. Given that mock data was drawn from an analytic function rather than one generated from the GP prior, these small discrepancies are not unexpected.

\section{Results} \label{sec:Results}

\begin{table}
    \centering
    \caption{Optimal hyperparameters from our sparse GP analysis of {\it Gaia} data. Units are ${\rm kpc}$ for $l_i$, ${\rm km\,s}^{-1}$ for $\sigma_f$ and ${\rm km\,s}^{-1}{\rm kpc}^{-2}$ for $\mu_i$ and $\nu_i$.}
    \label{tab:GPhyp}
    \begin{tabular}{llll}
        \hline
       & $V_R$ & $V_\phi$ & $V_Z$ \\
        \hline
        Kernel & RBF & RBF  $+$  lin2  & RBF\\
         $l_x$ & 1.06 & 1.13 & 1.89\\
        $l_y$ & 2.72 & 1.75 & 4.18 \\
         $l_z$ & 0.536 & 0.242 & 0.484\\
        $\sigma_f$ & 12.4 & 10.4 & 3.31 \\
        $\mu_x,\,\nu_x$ & & 0.236,~~0.228 & \\
        $\mu_y,\,\nu_y$ & & 0.207,~~0.209 & \\
        $\mu_z,\,\nu_z$ & & 3.86,~~3.71 &\\
    \end{tabular}
\end{table}

As with the mock data, we analyse the {\it Gaia} DR2 sample by first determining the mean values and uncertainties for the three velocity components in our $27k$ cells. Hyperparameters from our optimization of the likelihood function are given in Table \ref{tab:GPhyp}.
We see that $l_y/l_x\sim 1.5-2.5$ and $l_x/l_z\sim 2-5$. The hierarchy of scales is expected given that the radial length scale of the disk is a factor of $5-10$ times larger than the scale height and that variations in the disk tend to be stronger in the radial direction than the azimuthal direction. 

\subsection{Velocity field in components}\label{sec:Bulk-Components}

\begin{figure}
	\includegraphics[width=\columnwidth]{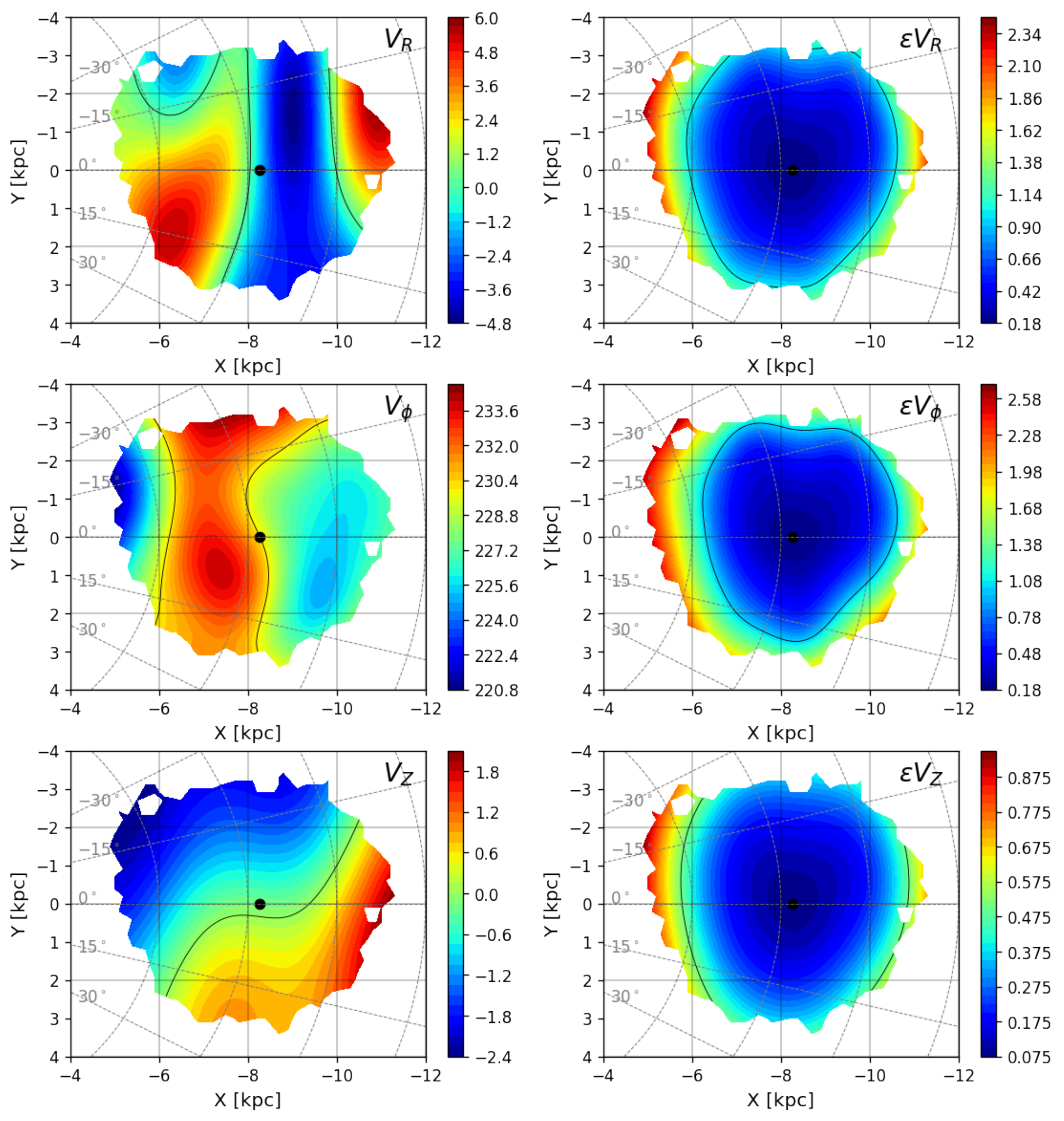}
    \caption{Left: Components of the velocity field $V_R$, $V_\phi$, and $V_z$ in the $z=0$ plane. Right: One-sigma confidence regions of the GP models.}
    \label{fig:VelocityFieldsXY}
\end{figure}

\begin{figure}
	\includegraphics[width=\columnwidth]{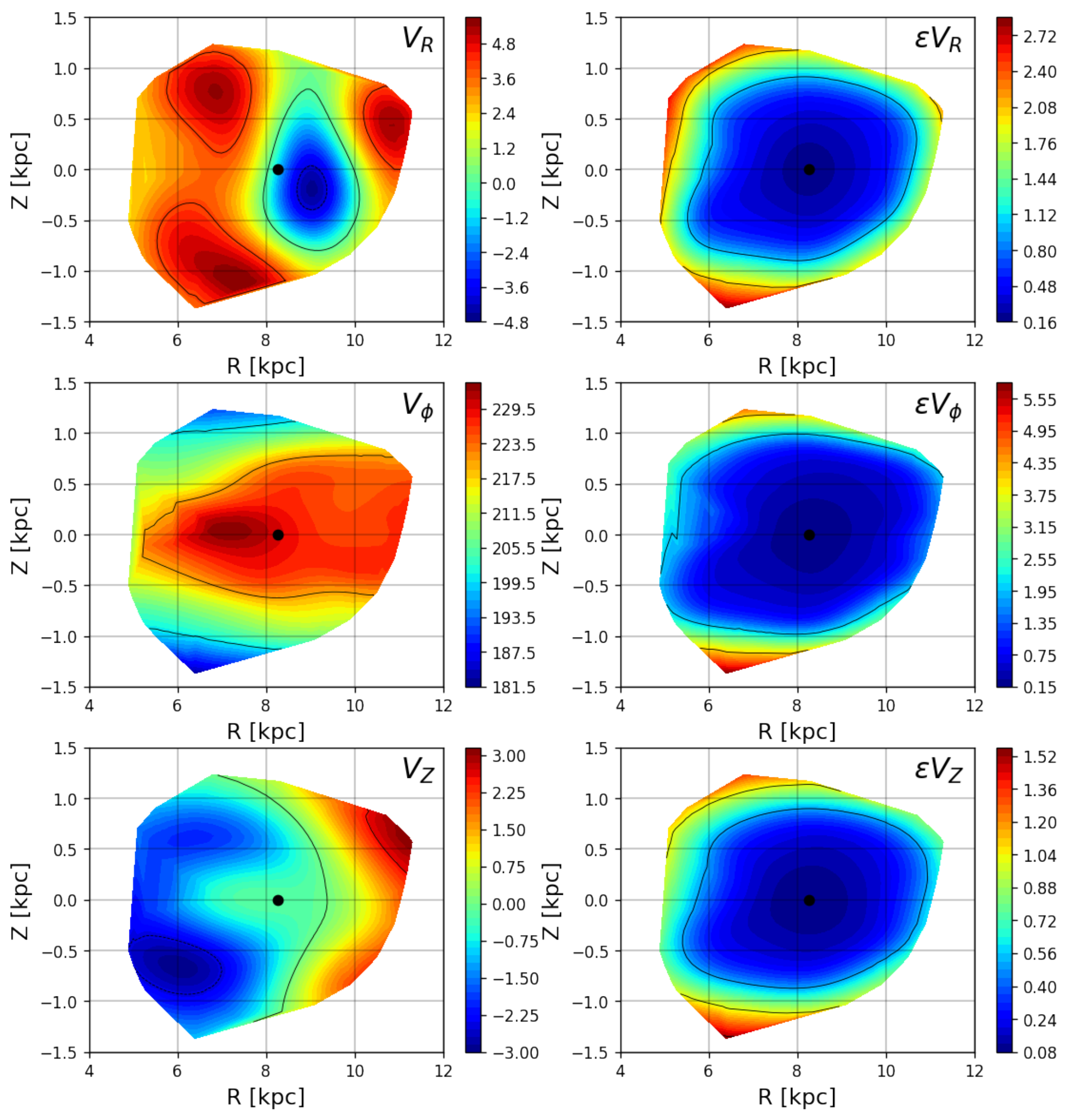}
    \caption{Left: Components of the velocity field in the $\phi=0$ plane. Right: One-sigma confidence regions of the GP models.}
    \label{fig:VelocityFieldsRZ}
\end{figure}

Figures \ref{fig:VectorfieldXY} and \ref{fig:VectorfieldRZ} show model predictions for the velocity field in the $z=0$ and $\phi=0^\circ$ planes. These figures are produced by querying the model at the appropriate points using Equation \ref{eq:posterior}. By and large, the maps are in good agreement with those found in \citet{Katz2019} (See their Figures 10 and 11) though they are clearly much smoother. The GP analysis tends to pick out features on the length scales characterized by $l_x$, $l_y$, and $l_z$. We also show the statistical uncertainties in the mean for the three components, $\epsilon_i$, as calculated from Equation \ref{eq:posteriorcov}. In general the uncertainties are less than $1\,{\rm km\,s}^{-1}$ within the sample volume, but rise rapidly as one approaches the edge of the volume. Our maps can also be compared to those in \citet{khanna2022} who combine data from {\it Gaia} EDR3 with radial velocity measurements from a number of spectroscopic surveys. They construct parametric models of the velocity field in heliocentric coordinates so a direct comparison is difficult.

The upper left panels of Figures \ref{fig:VelocityFieldsXY} and \ref{fig:VelocityFieldsRZ} point to several regions of radial bulk flows. In particular, there is inward bulk motion just beyond the Solar circle that is approximately independent of $\phi$ across the sample volume. This motion is primarily in the region $|z|<500\,{\rm pc}$ and is thus likely associated with the thin disk. On the other hand, the outward radial flow just inside the Solar circle is mainly found at positive $\phi$ but across the full range in $z$ of the survey. There also appears to be another outward flow at $R>10\,{\rm kpc}$, $\phi\simeq -10^\circ$ and positive $z$.

The middle-left panels of Figures \ref{fig:VelocityFieldsXY} and \ref{fig:VelocityFieldsRZ} show $V_\phi$. The dominant flow here, and indeed for ${\bf V}$ in general, is the motion about the Galactic centre. We see that contours of constant $V_\phi$ roughly follow contours of constant $R$ in Figure \ref{fig:VectorfieldXY}. As discussed below, the average of $V_\phi$ over $\phi$ at $z=0$ yields the rotation curve near the Solar circle. The main feature in Figure \ref{fig:VelocityFieldsRZ} is a decrease of $V_\phi$ as one moves away from the midplane of the Galaxy. This trend is easily explained by an increase in asymmetric drift due to a larger contribution from dynamically warmer stars in the thick disk. As with $V_R$, we see that $V_\phi$ is not perfectly symmetric about the midplane. Consider, for example, the ridge in $V_\phi$ between $X = -6$ and $X= -8$ in Figure \ref{fig:VectorfieldXY}, which coincides with the peak seen in Figure \ref{fig:VectorfieldRZ} at $R=7\,{\rm kpc}$. The peak and its outer slopes are shifted by a small, but non-negligible, amount above the midplane. 

The bottom left panels of Figures \ref{fig:VelocityFieldsXY} and \ref{fig:VelocityFieldsRZ} show our results for the vertical bulk motion, $V_Z$. In the midplane we see a clear trend of increasing $V_Z$ as one moves across the local patch of the Galaxy in the direction of increasing $R$ and $\phi$. The view in the $Rz$ plane shows a mix of bending and breathing motions. The disk is moving downward inside the Solar circle and upward outside the Solar circle and in addition it appears to be expanding away from the midplane. The generation of bending and breathing waves through internal disk dynamics and interactions with satellites such as the Sagittarius dwarf have been studied in N-body simulations by numerous authors including \citep{gomez2013,Chequers2018,laporte2019, poggio2021a, bennett2022} and \citet{thulasidharan2022}. Observational studies using data from various surveys in \citet{widrow2012, Williams_2013, carlin2013, carrillo2018, Katz2019, Wang2020, lopez2020}.

\subsection{Velocity vectors} \label{sec:bulk-vector}

\begin{figure}
	\includegraphics[width=\columnwidth]{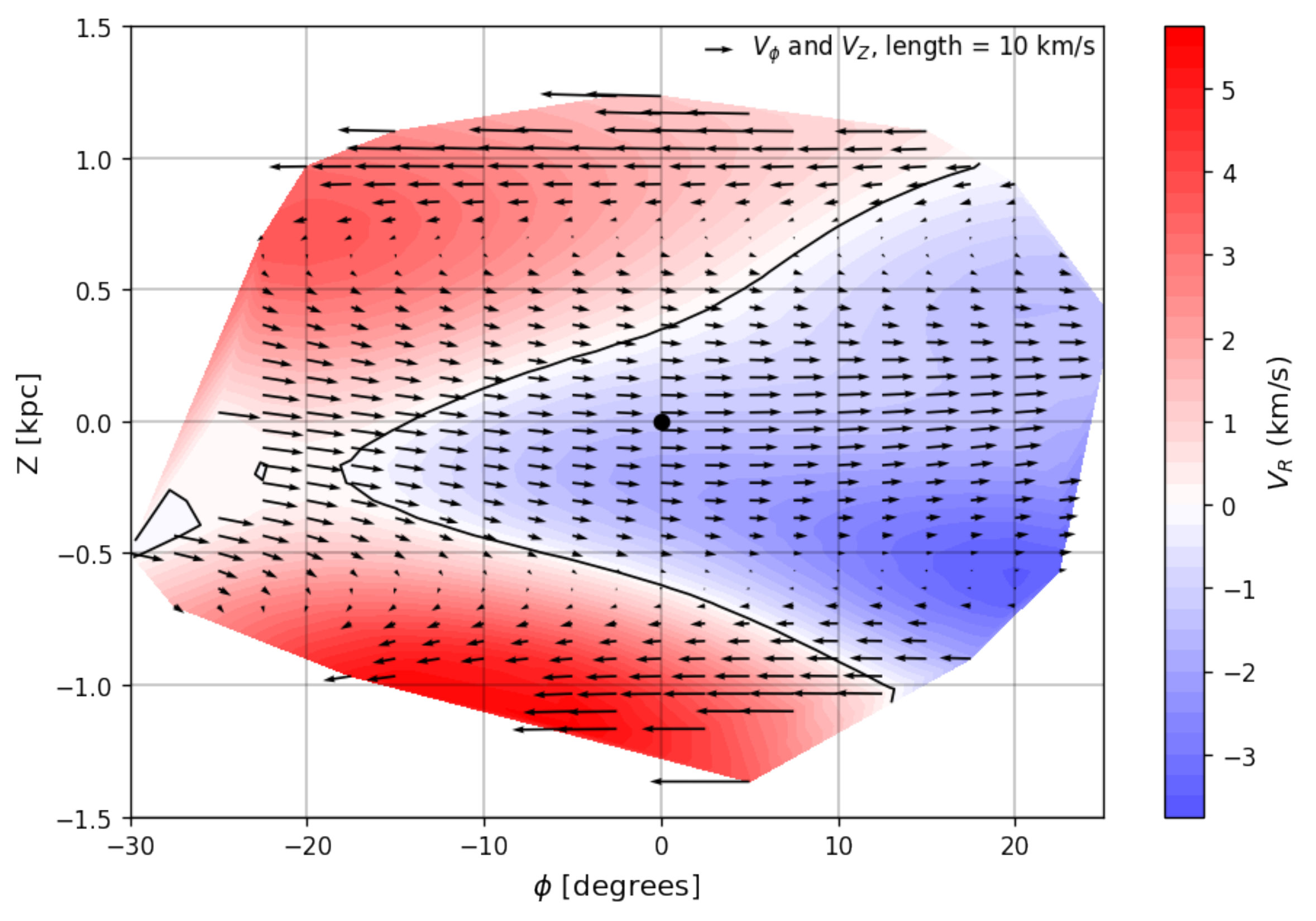}
    \caption{Map of the velocity field in $\phi-Z$ plane along a cylindrical surface at $R=8.27$. Arrows represent the components $V_\phi$ and $V_Z$ where, for clarity, we subtract $220\,{\rm km\,s}^{-1}$ from $V_\phi$. The background color map shows $V_R$ while black contour indicates the curve where $V_R=0$.}
    \label{fig:VectorfieldPhiZ}
\end{figure}

To aid our understanding of the velocity field we present velocity vector maps in three projections. Similar vector field maps were presented  in \citet{pearl2017} using data from LAMOST and, most recently, by \citet{Fedorov2022} using data from {\it Gaia} DR2. We begin with Figure \ref{fig:VectorfieldPhiZ}, which presents the vector field in the $\phi Z$ plane at $R=R_0$, that is, along a curved cylindrical surface that includes a part of the Solar circle. To help visualize the velocity field, we have subtracted off the vector $220$ \kms $\hat{\mathbf{\phi}}$. Radial velocities are shown as a color map. Thus, all three components of the velocity field are represented in the figure.

\begin{figure}
	\includegraphics[width=\columnwidth]{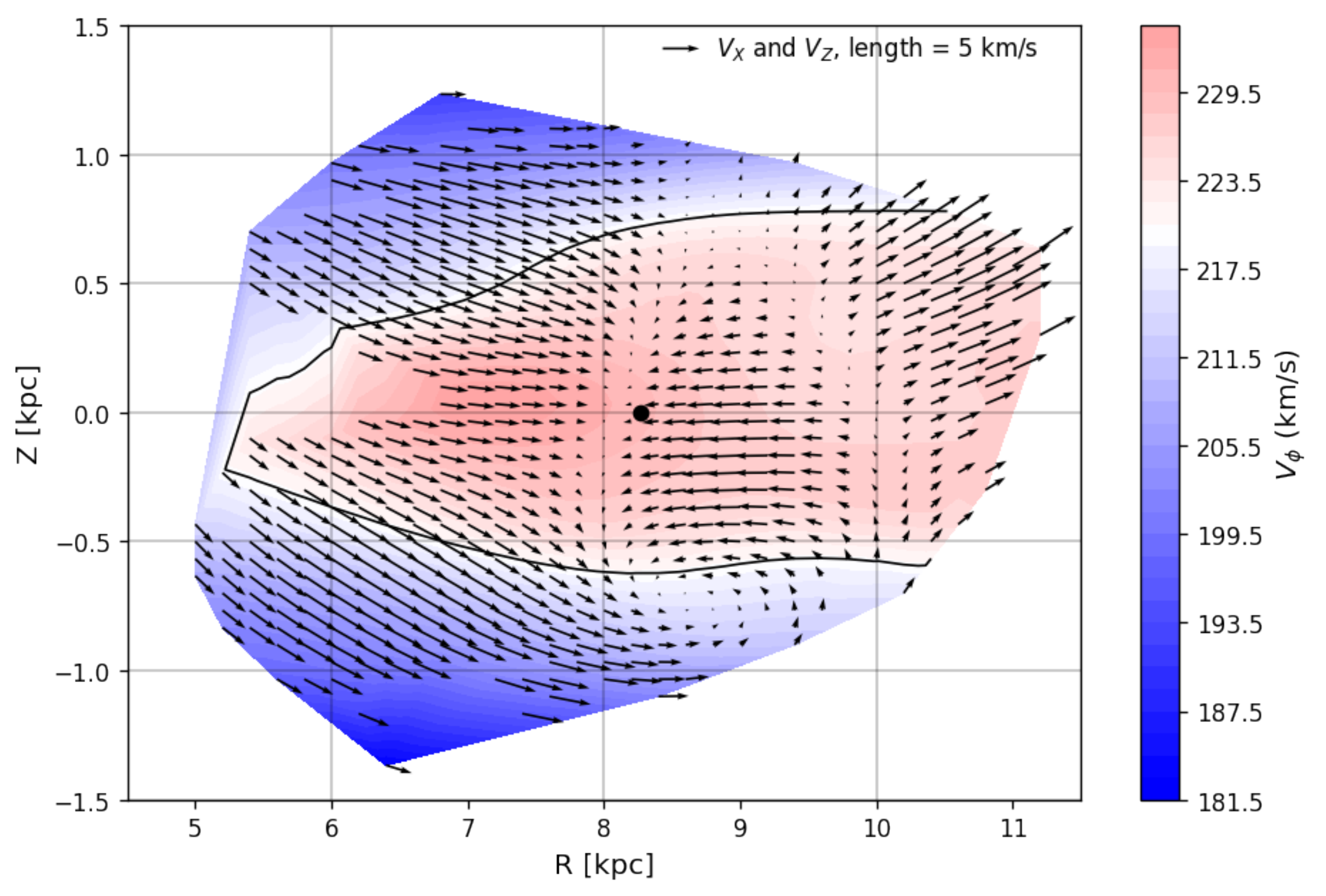}
    \caption{Velocity vector map in the $RZ$-plane for $\phi=0^\circ$ with arrows representing $V_R$ and $V_Z$. The background color map shows
$V_\phi$ with the black contour indicating the curve where $V_\phi = 220$ \kms. }
    \label{fig:VectorfieldRZ}
\end{figure}

As already noted in the $V_\phi$ panel of Figure \ref{fig:VectorfieldRZ}, the dominant feature of the map is the decrease in $V_\phi$ due to asymmetric drift as one moves away from the midplane. Velocities in the midplane are about $10\,{\rm km\,s}^{-1}$ higher than they are at $|Z| = 500\,{\rm pc}$. Though the dominant flow is in the azimuthal direction, we do see a clear downward motion for $\phi<0^\circ$, in agreement with Figure \ref{fig:VectorfieldXY}.

In Figure \ref{fig:VectorfieldRZ} we show the velocity field in the $RZ$-plane at $\phi=0^\circ$, which passes through the position of the Sun. We see that there are three distinct regions defined primarily by the sign of $V_R$. Specifically, we find an inward flow centered on $(R,\,z)=(8.5,-0.2){\rm kpc}$, an outward and downward flow inside the Solar circle, and an outward and upward flow for $R>10\,{\rm kpc}$ and $z\sim 500\,{\rm pc}$. The results are in good agreement with Figure 10 of \citet{pearl2017} for regions where the samples overlap.

\subsubsection{XY vectorfield} \label{sec:XY-vector}

\begin{figure}  
	\includegraphics[width=\columnwidth]{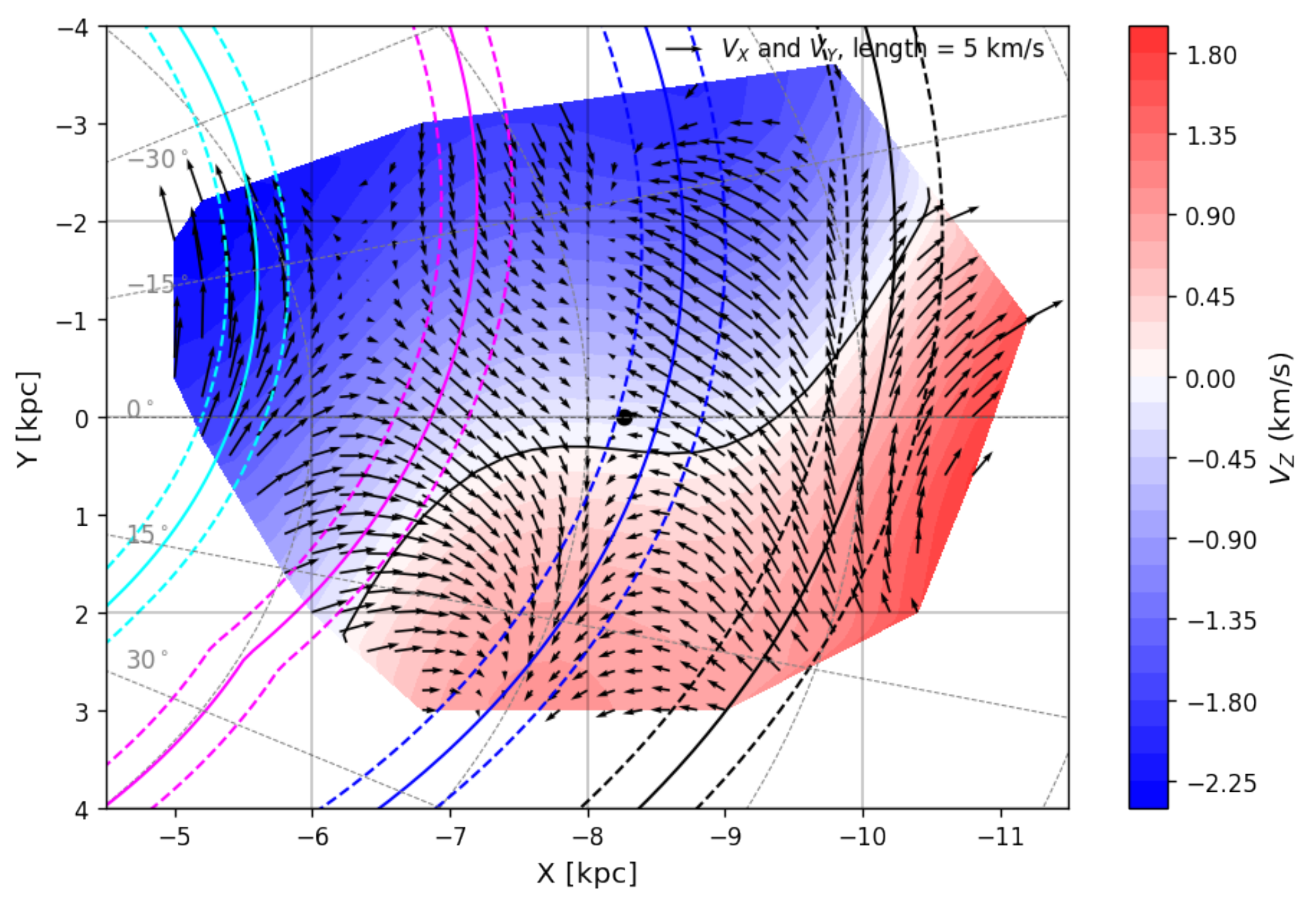}
    \caption{Velocity vector map in midplane. Arrows represent the $V_X$ and $V_Y$ components with the vector $230\,{\rm km\,s}^{-1}\hat{\mathbf{\phi}}$ subtracted for clarity.
    The background colormap shows $V_Z$ and the black contour indicates the curve where $V_Z = 0$. Arcs show the model for nearby spiral arms from \citet{Reid2019}. In order from the inner to the outer disc they are: Scutum (cyan), Sagittarius (magenta), Local Arm (blue), and Perseus (black).}
    \label{fig:VectorfieldXY}
\end{figure}

In Figure \ref{fig:VectorfieldXY} we show the velocity field in the midplane with vertical bulk motions are shown as the background color map. The map paints an image of several flows in the vicinity of the Sun, which can be describe by a combination of expansion or compression and shear. For example, roughly along the Solar circle, we find radial compression and azimuthal shear --- the dominant motion inside the Solar circle is outward and toward positive $\phi$ while the motion just beyond the Solar circle is inward and toward negative $\phi$.

It is tempting to attribute flows in the midplane to the presence of spiral arms. With this in mind, we overlay the spiral arm model of \citet{Reid2019}. In general, one expects to see motion toward the Galactic centre on the leading side of a spiral arm and motion away from the Galactic centre on the trailing side \citep{Kawata2014,Grand2015,Hunt2015}. This pattern is indeed seen along the Local Arm in Figure \ref{fig:VectorfieldXY} though the connection between flows in our map and the other arms is more tenuous. For example, there is little evidence for changes in the velocity field along the Sagittarius arm. This may be because Sagittarius is a relatively minor arm of the Galaxy and mainly a region rich in star formation but having relatively little mass \citep{Churchwell2009}. Analysis of density enhancement \citep{Benjamin2005} and Galactocentric rotation \citep{Kawata2018} show few of the expected features for the other arms.


\section{Velocity Gradient}\label{sec:Velocity-Gradient}

In the previous section, we showed that our GP model could be queried to
to give the bulk velocity field ${\bf V}$ at any position in the sample region. Since the GP solution is continuous and differentiable, we can also use it to infer the gradient of the velocity field, a $3\times 3$-tensor whose components are $\partial V_i/\partial X_j$. These components can be computed either by replacing the kernel functions in Equations \ref{eq:posterior} and \ref{eq:posteriorcov} with their gradients or by a finite difference scheme. We have confirmed that the two methods agree for the RBF kernel. In what follows we use finite differencing. Similar results are presented in \citet{Fedorov2021}.

\subsection{Oort constants}
Gradients of the velocity field played a central role in Oort's seminal work on the rotation of the Galaxy \citet{Oort1927a}. The basic idea was to expand the velocity field near the Sun as a Taylor series in position. Though Oort considered only axisymmetric flows in the $X-Y$ plane, the method was extended to general three dimensional flows by
 \citet{Ogrodnikoff1932,Milne1935,Chandrasekhar1942} and \citet{Ogorodnikov1965}.
 
To linear order in position, the velocity field near the Sun can be written as a Taylor series
\begin{equation} \label{eq:expansion}
 {\bf V} = {\bf V}_\odot + \mathbf{H}\cdot\B{x} + {\cal O}(\B{x}^2)~.
\end{equation}
If we restrict ourselves to the projection of the velocity field onto the midplane of the Galaxy, then ${\mathbf{H}}$ reduces to the $2\times 2$ matrix
\begin{equation} \label{eq:Oort-cart}
  \mathbf{H} = \left(
    \begin{array}{c@{\;\;}c}
      \partial v_x/\partial x & \partial v_y/\partial x \\ 
      \partial v_x/\partial y & \partial v_y/\partial y
    \end{array} 
  \right)_{{\bf x}=0} \equiv \left(
    \begin{array}{c@{\;\;}c}
      K+C & A-B \\
      A+B & K-C
    \end{array}
  \right).
\end{equation}
where the second equality defines the Oort constants
$A$, $B$, $C$, and $K$. These constants measure, respectively, the azimuthal shear, vorticity, radial shear, and divergence of the velocity field in the midplane of the disk. In Galactocentric polar coordinates $(R,\varphi)$ the Oort constants are given by
\begin{eqnarray} 
    \label{eq:Oort-Cyl-1}
    2A &=&  \frac{v_\varphi}{R} - \frac{\partial v_\varphi}{\partial R} - \frac{1}{R} \frac{\partial v_R}{\partial\varphi} \\
    \label{eq:Oort-Cyl-2}
    2B &=&   -  \frac{v_\varphi}{R} - \frac{\partial v_\varphi}{\partial R} +  \frac{1}{R} \frac{\partial v_R}{\partial\varphi}\\
    \label{eq:Oort-Cyl-3}
    2C &=&   -  \frac{v_R}{R} + \frac{\partial v_R}{\partial R}-  \frac{1}{R} \frac{\partial v_\varphi}{\partial\varphi}\\
    \label{eq:Oort-Cyl-4}
    2K &=& \frac{v_R}{R} + \frac{\partial v_R}{\partial R}+ \frac{1}{R} \frac{\partial v_\varphi}{\partial\varphi}
\end{eqnarray}
\citet{Chandrasekhar1942}. 

The usual method for determining the Oort constants from astrometric data is based on the observation that the proper motion of a star in the direction of Galactic longitude, $\mu_l$, and the line-of-sight velocity, $v_{\rm los}$, can be expressed in terms of sine and cosine functions of $l$. The Oort constants appear as coefficients in this truncated Fourier series and can therefore be determined by standard statistical methods. (See, for example, \citet{Feast1997,Olling2003,Bovy2017,Vityazev2018,Wang2021}

\begin{table}
    \centering
    \caption{Comparison of measured values for the Oort constants from the literature and from this work in units of \kmskpc. For the GP model we provide the inferred values at the exact position of the Sun and average values over a spherical volume of radius $500\,{\rm pc}$.}
    \label{tab:Oort-Consts}
    \begin{tabular}{lllll}
        \hline
       Origin & $A$ & $B$ & $C$ & $K$  \\
        \hline
        Oort (\citeyear{Oort1927a}) & \raisebox{0.7ex}{\texttildelow}19 & \raisebox{0.7ex}{\texttildelow}-24  & &\\
        Feast (\citeyear{Feast1997}) & 14.8 $\pm$ 0.8 & -12.4 $\pm$ 0.6 & &\\
        Olling (\citeyear{Olling2003})& 15.9 $\pm$ 2 & -16.9 $\pm$ 2 & -9.8 $\pm$ 2 & \\
        Bovy (\citeyear{Bovy2017}) & 15.3 $\pm$ 0.4 & -11.9 $\pm$ 0.4 & -3.2 $\pm$ 0.4 & -3.3 $\pm$ 0.6 \\
        Vityazev (\citeyear{Vityazev2018}) & 16.3 $\pm$ 0.1 & -11.9 $\pm$ 0.1 & -3.0 $\pm$ 0.1 & -4.0 $\pm$ 0.2 \\
        Li (\citeyear{Li2019}) & 15.1 $\pm$ 0.1 & -13.4 $\pm$ 0.1 & -2.7 $\pm$ 0.1 & -1.7 $\pm$ 0.2\\
        Wang (\citeyear{Wang2021}) & 16.3 $\pm$ 0.9 & -12.0 $\pm$ 0.8 & -3.1 $\pm$ 0.5 & -1.3 $\pm$ 1.0\\
        This work (p) & 16.2 $\pm$ 0.2 & -11.7 $\pm$ 0.2 & -3.1 $\pm$ 0.2 & -3.0 $\pm$ 0.2 \\
        This work (v) & 15.2 $\pm$ 0.8 & -12.4 $\pm$ 0.9 & -2.9 $\pm$ 0.8 & -2.3 $\pm$ 0.5 \\
    \end{tabular}
\end{table}

In this paper, we compute the Oort constants directly from our GP model for the velocity field using the above equations. Our results along with a selection of values from the literature are given in Table \ref{tab:Oort-Consts}. We find excellent agreement with recent measurements from \citet{Bovy2017,Vityazev2018,Li2019} and \citet{Wang2021}. As noted above, the values from the literature all determine the Oort constants by fitting $\mu_l$ and $v_{\rm los}$ to a low-order Fourier series in $l$. They do, however, use data from different surveys and with different geometric selection functions and sample sizes. For example, both \citet{Bovy2017} and \citet{Li2019} 
consider large samples of stars within $500\,{\rm pc}$ of the Sun. \citet{Bovy2017} considers main sequence stars from the Tycho–{\it Gaia} Astrometric (TGAS) catalog \citep{TGAS2015} while \citet{Li2019} considers all stars from {\it Gaia} DR2. \citet{Wang2021} uses a relatively small sample of $A$-stars from LAMOST with a similar range in distance from the Sun. Finally, \citet{Vityazev2018} uses stars from TGAS but with the larger reach of $1.5\,{\rm kpc}$.

Our model allows us to predict the Oort constants at a single point, namely the position of the Sun. These values are given in the second to last line in Table \ref{tab:Oort-Consts}. To allow for a closer comparison to literature values, we also include values for the Oort constants averaged over a spherical volume of radius $500\,{\rm pc}$.

\subsection{Oort functions}\label{sec:Oort-functions}

\begin{figure}
    \centering
    \includegraphics[width=\columnwidth]{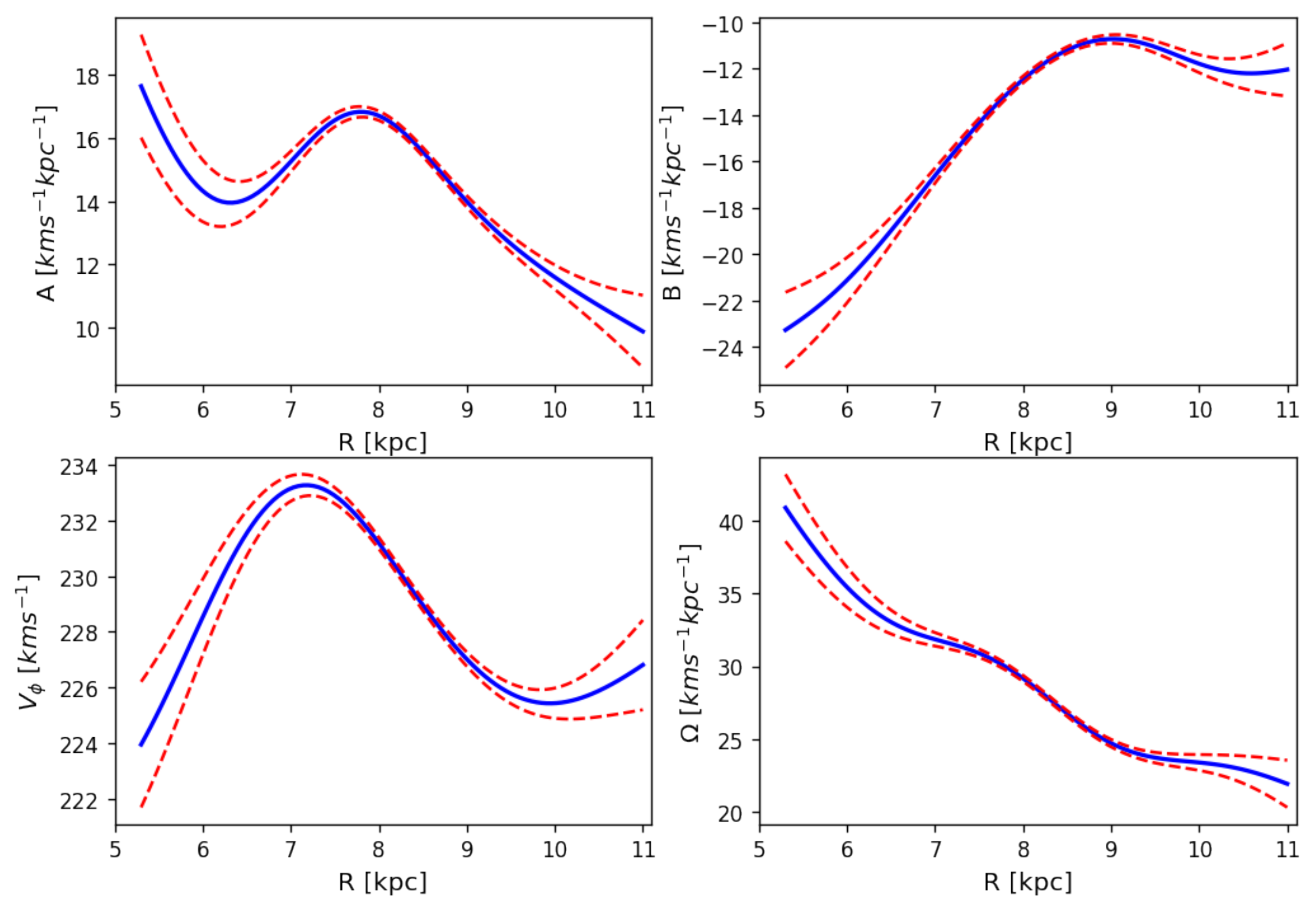}
    \caption{Galactic parameters as a function of radius. Top: Oort parameters $A$ (left) and $B$ (right). Bottom left: $V_\phi$. Bottom right: Angular velocity $\Omega = A - B$. The model is queried along the line passing from the Galactic centre through the Sun. The red dashed lines show the $1\,\sigma$ uncertainties.}
    \label{fig:Oort-functions}
\end{figure}

\begin{figure}
	\includegraphics[width=\columnwidth]{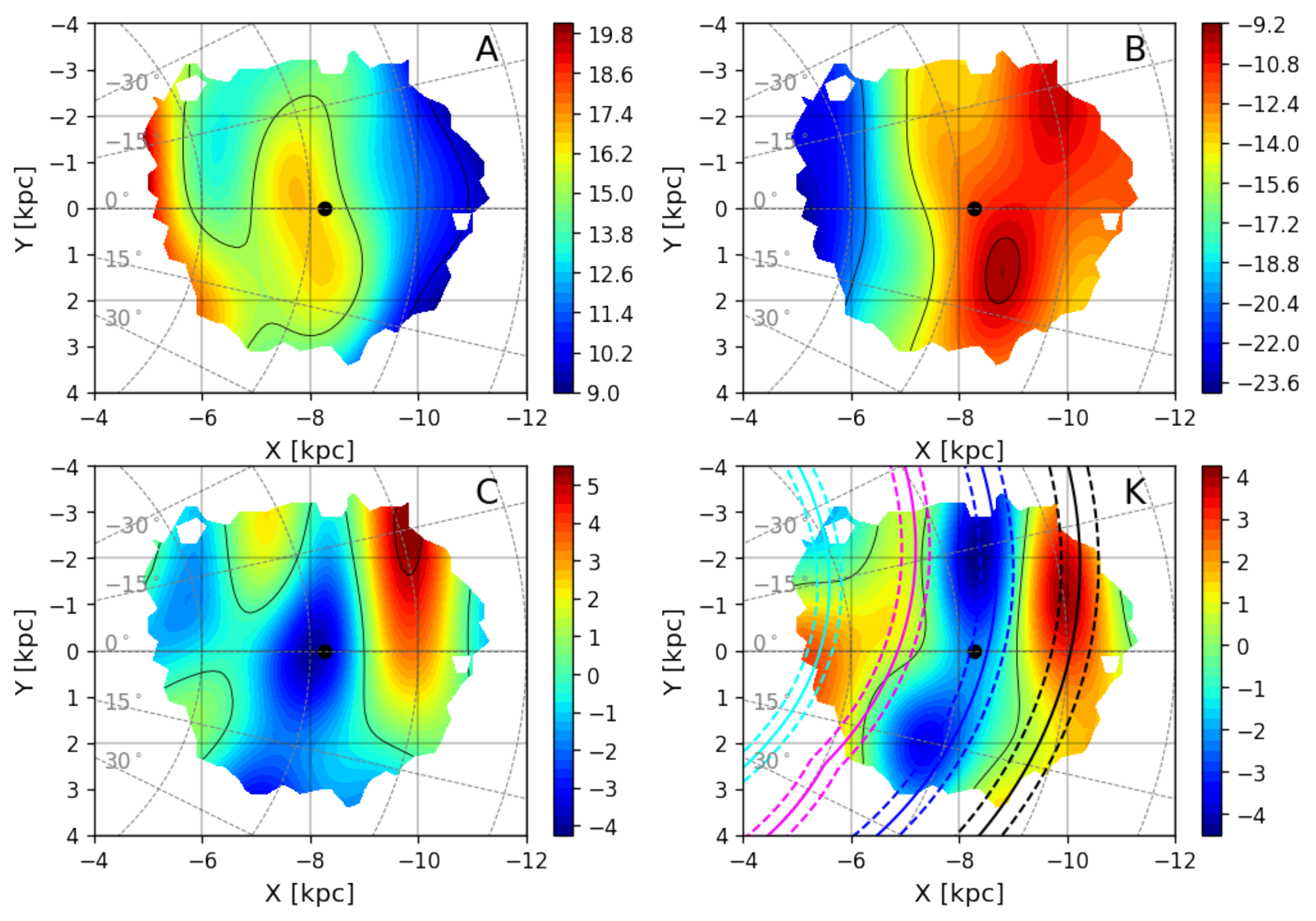}
    \caption{Oort constants $A$, $B$, $C$, and $K$ calculated from Equations \ref{eq:Oort-Cyl-1} - \ref{eq:Oort-Cyl-4}. Colormap units are in \kmskpc. The spiral arms model of \citet{Reid2019} is also over-plotted for the local divergence ($K$). In order of inner to the outer disc the arms are: Scutum (cyan), Sagittarius (magenta), Local Arm (blue), and Perseus (black).}
    \label{fig:Oort-Consts}
\end{figure}

In Oort's \citeyear{Oort1927a} original work, stars are assumed to follow circular orbits in the midplane of the Galaxy. Under this assumption, $C$ and $K$ are identically zero and $A$ and $B$ can be written in terms of circular speed and its gradient at the position of the Sun. They therefore provide a direct probe of the gravitational potential and hence matter distribution in the Galaxy. \citet{Olling1998} extended the idea of Oort constants to Oort functions with 
\begin{eqnarray}
    A(R) & = \frac{1}{2} \left (\frac{v_c(R)}{R} - \frac{dv_c}{dR}\right )\\
        B(R) & = \frac{-1}{2} \left (\frac{v_c(R)}{R} + \frac{dv_c}{dR}\right )
\end{eqnarray}
where $v_c$ is the circular speed. They then derived $A(R)$ and $B(R)$ from mass models of the Milky Way and Oort constant measurements. 

In Figure \ref{fig:Oort-functions} we show $A(R)$ and $B(R)$ over the sample region. Note that our results for $A(R)$ and $B(R)$ use mean $V_\phi$ rather than $v_c$ and therefore include the effects of asymmetric drift. Nevertheless, they show the same broad trends as seen in \citet{Olling1998} and more recently, \citet{Fedorov2021}. $A(R)$ is a decreasing function of $R$ with a "bump" just inside the Solar circle. $B(R)$ is an increasing function of $R$. In the bottom panels of Figure \ref{fig:Oort-functions} we show the rotation curve $V_\phi$ and 
angular velocity curve $\Omega = |A-B|$, which show departures from a flat rotation curve at the few ${\rm km\,s}^{-1}$ level.

The expressions for the Oort constants in Equations \ref{eq:Oort-Cyl-1}-\ref{eq:Oort-Cyl-4} can similarly be extended to function of $R$ and $\varphi$ to yield maps of Oort functions in the Galactic midplane. These are shown in Figure \ref{fig:Oort-Consts}. Were the Galaxy in an axisymmetric steady state, we would find that $C=K=0$ and $A$ and $B$ depend only on $R$. Though $C$ and $K$ are about a factor of five smaller than $A$ and $B$ they are clearly nonzero. Moreover, all four functions show variations in $\varphi$ though the dominant gradients are in the $R$-direction. The prominent regions of shear (lower left panel showing $C$) and compression (lower right panel showing $K$) provide another way of visualizing the in-plane flows already seen in Figure \ref{fig:VectorfieldXY}.

\subsection{Divergence of the velocity field}\label{sec:Divergence}
\begin{figure}
	\includegraphics[width=\columnwidth]{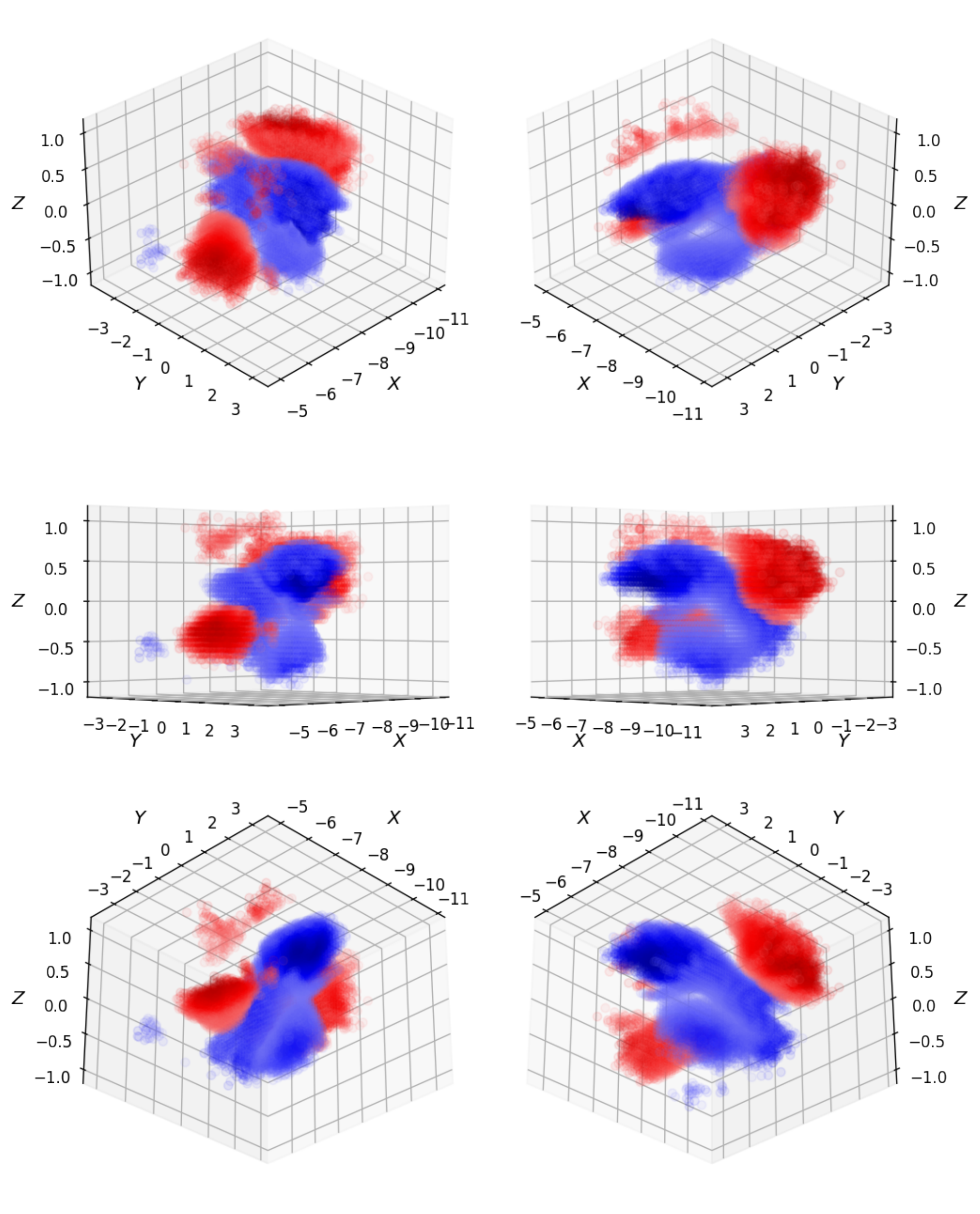}
    \caption{Divergence of the velocity field, calculated using cylindrical coordinates. Regions with $|\nabla \cdot \mathbf{V}| > 5$ \kmskpc are plotted, positive in red and negative in blue. Coloring and opacity is scaled with amplitude of the divergence. }
    \label{fig:Divergence}
\end{figure}

As mentioned above, the Oort constants are constructed from $X-Y$ derivatives of the in-plane components of the velocity field. More generally, ${\bf H}$ in Equation \ref{eq:expansion} is a $3\times 3$ tensor. This tensor can be written as the sum of a six-component symmetric tensor $M^+$ and a three-component anti-symmetric tensor $M^-$
\begin{equation}\label{eq:O-M_model}
    M^\pm = \frac{1}{2}\left (\frac{\partial v_i}{\partial x_k} \pm \frac{\partial v_k}{\partial x_i}\right )
\end{equation}
where as before, the components are evaluated at the position of the Sun
\citep{Ogrodnikoff1932, Milne1935,Ogorodnikov1965, Tsvetkov2019, Fedorov2021}. Note that the coefficients of $M^\pm$ include the Oort constants. For example, $A = M^+_{12}$ and $B = M^-_{21}$. Recently, the nine parameters of the velocity gradient method have been estimated using data from {\it Gaia} DR2 \citep{Vityazev2018, bobylev2021}.

Here, we focus on the divergence of the velocity field $\mathbf{\nabla}\cdot {\bf V}$, which corresponds to the trace of $M^+$. Recall that the zeroth moment of the collisionless Boltzmann equation yields the continuity equation
\begin{equation}
    \frac{\partial \rho}{\partial t} + \mathbf{\nabla}\cdot 
    \left (\rho {\bf V}\right ) = 0
\end{equation}
\noindent where $\rho$ is the stellar density. This equation may be written as
\begin{equation}
\frac{1}{\rho}\frac{d\rho}{dt} = -\mathbf{\nabla}\cdot {\bf V}
\end{equation}
where $d/dt = \partial/\partial t + {\bf V}\cdot \mathbf{V}$ is the total derivative. Thus, a region of positive divergence indicates that either the stellar distribution is expanding, or stars are being transported into a region of lower density. Either way, $\mathbf{\nabla}\cdot {\bf V}$ is a clear signal of disequilibrium in the disk.

In Figure \ref{fig:Divergence} we show the divergence in three dimensions from six different viewing angles. We find a prominent region of negative divergence corresponding to compression in the velocity field running through the Solar Neighborhood and roughly aligned with Galactic azimuth. This is in good agreement with the results for $K$ in the Galactic plane as seen in the lower-right panel of Figure \ref{fig:Oort-Consts}. This region is bracketed by regions of positive divergence, one below the midplane and inside the Solar circle and the other above the midplane and outside the Solar circle. The amplitude of the divergence is $5-10 \kmskpc$, which implies that the time-scale for the logarithm of the density to change is of order $100-200\,{\rm Myr}$.

\section{Discussion}\label{sec:Discussion}

Our discussion of the divergence of the stellar velocity field illustrates how a GP model can provide a link between theory and data. In particular, the divergence gives us a direct indication of regions in the disk where the {\it total} time derivative of $\ln{\rho}$ is nonzero. Clearly, if we have $\rho$ itself, then we can estimate the convective derivative term ${\bf V}\cdot \mathbf{\nabla}\rho$ and hence infer the {\it partial} time derivative of $\rho$. In this way, we gain information about time-dependent phenomena in the disk from a kinematic snapshot. Models of $\rho$ for particular tracer populations require detailed knowledge of the selection function but are certainly accessible given the wealth of astrometric data available.

In a similar fashion, GP models for components of the velocity dispersion tensor would allow one to study the first moments of the CBE, namely the Jeans equations. In contrast with the continuity equation, these equations involve gradients of the gravitational potential. Thus, we will typically have two unknown quantities, one involving time-derivatives of the moments and the other involving the potential. Nevertheless, one should be able to infer something about the dynamics of the stellar disk, and in particular, departures from equilibrium by modelling moments of the DF via GP regression.

These considerations suggest an extension of the present work where we treat the output as a single three-components vector rather than three independent scalars. Doing so would allow for correlations between the different components of the velocity field. The framework for extending GP regression to vector outputs is already well-established \citep{alvarez2011}.

\section{Conclusions}\label{sec:Conclusion}

In this paper, we present a GP model for the mean or bulk stellar velocity field in the vicinity of the Sun using astrometric measurements from {\it Gaia DR2}. The model is nonparametric in the sense that there are no prior constraints on the functional form of the velocity field. Instead, one specifies the functional form of the kernel function, which through a set of hyperparameters, controls properties of the prior on the velocity field such as its coherence length. 

The main challenge in applying GP regression to {\it Gaia} data comes from the large-$N$ requirements in both computing time and RAM. Fortunately, there is a large effort within the machine learning community on addressing these problems (See, for example, \citet{hensman2013}). In this work, we have just started to exploit methods developed in that field.

Our model provides smooth, differentiable versions of the velocity field maps found in \citet{Katz2019} and elsewhere. The {\it Gaia} maps used different binning schemes to compute different projections of the velocity field. In our case, we compute a single GP model (or more precisely, three independent models for each of the cylindrical velocity field components) from which properties of the velocity field could be derived. As a check of the model, we confirm that the values for the Oort constants derived by differentiating the velocity field agree extremely well with recent determinations in the literature.

A GP model for the velocity field can thus provide a starting point for making contact between astrometric data and dynamics via the continuity and Jeans equations. Already, our results for the divergence of the velocity field have provided a unique perspective on departures in the disk from an equilibrium state.

\section*{Acknowledgements}

It is a pleasure to thank Dan Foreman-Mackey, Hamish Silverwood, and Lauren Anderson for helpful conversations on Gaussian processes. We acknowledge the financial support of the Natural Sciences and Engineering Research Council of Canada. LMW is grateful to the Kavli Institute for Theoretical Physics at the University of California, Santa Barbara for providing a stimulating environment during a 2019 program on galactic dynamics. His research at the KITP was supported by the National Science Foundation under Grant No. NSF PHY-1748958.

\section*{Data Availability}

Data used in this paper is available through Zenodo (https://doi.org/10.5281/zenodo.2557803)



\bibliographystyle{mnras}
\bibliography{NelsonWidrow} 








\bsp	
\label{lastpage}
\end{document}